\documentstyle[12pt]{article}
\footskip- 15mm
\begin{document}

\begin{center}

Chaos control in  photoconductors\\

ELMAN M. SHAHVERDIEV,\footnote{e-mail:shahverdiev@lan.ab.az}\\

Institute of Physics of Academy of Sciences of Azerbaijan,
33,H.Javid Av., 370143-Baku,Azerbaijan.\\
\end{center}
1.Introduction.\\
Nowadays, it is well established that the chaotic behavior is the 
intrinsic property of some of the nonlinear dynamical systems [1-6]. 
Due to the nonlinearity most of the rate equations (charge-carrier 
dynamics) in semiconductor physics, such an unpredictable behavior 
also can be easily detected in most of the realistic models of 
photoconductors. The technological importance of photoconductivity, 
particularly its applications in high-speed and far-infrared 
photodetectors and similar devices, leads to the need for a careful 
investigation of the charge-carrier dynamics. One of such models was 
proposed recently in [7], see also [8-9]. To be more specific, that 
model consists of three coupled nonlinear ordinary differential 
equations describing the population densities of electrons $n$, 
trapped electrons $m$, and holes $p$ in the following form of rate 
equations
$$\frac{dn}{dt}=G-n\alpha_{1} (N_{t}-m)+\gamma_{1} m-c_{1} n,$$
$$\frac{dm}{dt}=n\alpha_{1}(N_{t}-m)-\delta_{0} mp$$
$$-\gamma_{1}m,\hspace*{3cm}(1)$$ 
$$\frac{dp}{dt}=G-\delta_{0}mp-c_{2}p,$$
where: $G$ is the rate of photoexcited charge carriers;$c_{1}$ and 
$c_{2}$ are the capture rates of active centers for electrons and 
holes; $N_{t}$ is the total number of traps in the system; $m$ is 
the number of occupied traps at time $t$; $\alpha_{1}$ is the 
probability of capture of the electrons from the conduction band to 
the traps; $\gamma_{1}$ is the probability of ejection of electrons 
from the traps to the conduction band by thermal 
excitation;$\delta_{0}$ is the recombination constant for captured 
electrons.
Let  
$$ n(t=0)=n_{1}, m(t=0)=m_{1}, p_(t=0)=p_{1},\hspace*{5cm}(2)$$
be the initial conditions.\\
In dealing with such systems (1) it is always convenient to use 
dimensionless systems: dimensionless variables, time and constants are
introduced by the following relationships:
$$x=nn_{0}^{-1}, y=mn_{0}^{-1}, z=pn_{0}^{-1},t_{1}=tt_{0}^{-1},$$
$$t_{0}=c_{1}^{-1}, g=Gt_{0}n_{0}^{-1},$$
$$ n_{0}=(\alpha_{1}t_{0})^{-1},$$
$$k=\alpha_{1} N_{t}t_{0}, \alpha =\gamma_{1}t_{0},$$
$$ \kappa =c_{2}t_{0},$$
$$\beta =\delta_{0} t_{0} n_{0}$$
$$=\delta_{0}\alpha_{1}^{-1},\hspace*{3cm}(3)$$
The corresponding dimensionless system is of the form:
$$\frac{dx}{dt_{1}}=g-(k+1)x+xy+\alpha y, $$
$$\frac{dy}{dt_{1}}=kx-xy-\beta yz-\alpha y, \hspace*{3cm}(4)$$
$$\frac{dz}{dt}=g-\beta yz- \kappa z, $$
According to [7-9], nonlinear system (4) exhibits chaotic behavior 
with some values of systems parameters $g, k$, $\beta$ , $\alpha$ 
,$\kappa$. Namely, using the data presented in [7]: 
$G=10^{24}cm^{-3}sec^{-1}$ (for the correct dimension purposes we use 
this value of $G$ instead of $G=10^{16}cm^{-2}sec^{-1}$ as in [7]); 
$N_{t}=5 10^{14}cm^{- 3}$;$\gamma_{1}=0.83 sec^{-1}$; 
$c_{1}=1.5 10^{-3}sec^{-1}$; $c_{2}=1.5 10^{-5}sec^{- 1}$;
$\alpha_{1}=2.5 10^{-15}cm^{-3}sec^{-1}$; 
$\delta_{0}=10^{-15}cm^{-3}sec^{-1}$
we obtain the following values for $t_{0}, n_{0}$: 
$t_{0}=\frac{10^{3}}{1.5}$,$n_{0}=6 10^{11}cm^{-3}$ and for 
dimensionless control parameter $g=1.1 10^{15}$ and other parameters 
for chaotic behavior to occur:$\beta =0.4$, $\kappa =10^{-2}$,
$k=8.33 10^{2}$, $\alpha =5.5 10^{2}$. As the chaotic, unpredictable 
behavior was uncontrolable, ungovernable, and so undesirable in some 
situations, initially researchers tried to avoid chaos and to deal 
with the systems in the range of the parameters, where chaos was not 
generated. The situation has changed dramatically since the discovery 
of the possibility of chaos control by Pecora and Carrol [10] , Ott, 
Grebogi and Yorke [11] in 1990. These seminal papers [10-11] have 
induced avalanche of research works in the field of chaos control. 
Chaos synchronization in dynamical systems is one of methods of 
controling chaos, see, e.g. [10-17] and references therein. The 
interest to chaos synchronization in part is due to the application 
of this phenomenen in secure communications, in modeling of brain 
activity and  recognition processes,etc [10-17]. Also it should be 
mentioned that this method of chaos control may result in improved 
performance of chaotic systems [10-17].In our case of charge-carrier 
dynamical system chaos control could be used for the optimization of 
the performance of photoconductor devices.\\
According to [10] synchronization of two systems occurs when the 
trajectories of one of the systems will converge to the same values as 
the other and they will remain in step with each other. For the 
chaotic systems synchronization is performed by the linking of 
chaotic systems with a common signal or signals (the so-called 
drivers): suppose that we have a chaotic dynamical system of 
three or more state variables. In the above mentioned way of chaos 
control one or some of these state variables can be used as an input 
to drive a subsystem consisting of remaining state variables and which 
is a replica of part of the original system.In [10] it has been shown 
that if the real parts of the Lyapunov exponents for the subsystem 
(below: sub-Lyapunov exponents) are negative then the subsystem 
synchronizes to the chaotic evolution of original system. If the 
largest sub-Lyapunov exponent is not negative, then one can use the 
nonreplica approach to chaos synchronization [15]. To be more 
specific, one can try to make negative the real parts of the 
conditional Lyapunov exponents of the nonreplica response system. As 
it has been shown in [15] from the application viewpoint nonreplica 
approach has some advantages over the replica one. The above-mentioned 
chaos synchronization method [10] (replica approach) is applied to 
different chaotic dynamical systems, see, [7-11] and references 
therein. As it is already underlined recently a new approach-
nonreplica approach to chaos synchronization is proposed in [15]. 
A detailed analysis of that paper shows that for high dimensional 
systems the calculation of conditional Lyapunov exponents in general 
requires formidable, tedious numerical and analytical calculations.\\
This paper is dedicated to the chaos synchronization in the charge-
carrier dynamical systems in photoconductors within both the replica 
and nonreplica approaches. It has been shown that by using the 
boundedness of the solutions to the dynamical systems, nonreplica 
approach to chaos synchronization and Routh-Hurwitz criteria it is 
possible to make negative all the conditional Lyapunov exponents 
without complex numerical and analytical calculations.This is the 
main feature of the paper.\\
Thus consider the possibility of chaos synchronization in the 
dynamical system (4). The system has the following steady state (fixed 
point) solutions:
$$ x^{st}=g(1+\beta \kappa ^{-1}y^{st})^{-1},$$ 
$$z^{st}=x^{st}\kappa^{-1},$$
$$\alpha \beta \kappa^{-1}(y^{st})^{2}+(\alpha +g$$
$$ +\beta\kappa^{-1}g)y^{st}-kg=0,\hspace*{2cm}(5) $$
For the begining we investigate system (1) qualitatively, as it has 
been shown in [7], for different values of parameters the system 
exhibits different types of solutions, such as oscillatory, chaotic, 
etc. After linearization about the steady state solutions (5) of the 
system (4) we obtain that characteristic equation's roots are 
satisfying the  following equation:
$$\lambda^{3} + a\lambda^{2} + b\lambda + c=0, \hspace*{5cm}(6)$$
where
$$a=k+1+x^{st}+(\beta -1)y^{st}+\alpha + \kappa +\beta z^{st},$$
$$c=(\kappa +\beta y^{st})(\alpha +x^{st})$$
$$+\kappa (k-y^{st}+1)\beta z^{st},$$
$$b=(k-y^{st}+1)(\kappa +\beta y^{st} +\beta z^{st})$$
$$+\beta \kappa z^{st}$$
$$+(\alpha +x^{st})(1+\kappa +\beta y^{st}),\hspace*{3cm}(7)$$
In order to have oscillating solutions, the value of 
$$F=\frac{c^{2}}{4}+\frac{b^{3}}{27}+\frac{ca^{3}}{27}$$
$$-\frac{abc}{6}-\frac{a^{2}b^{2}}{108},\hspace*{2cm}(14)$$
should be positive [18].The period of the oscillations could be 
estimated by the following relationship:
$$T=3^{-0.5}\pi ((-\frac{q}{2}+F^{0.5})^{\frac{1}{3}}$$
$$ - (-\frac{q}{2}$$
$$- F^{0.5})^{\frac{1}{3}})^{-1},\hspace*{0.3cm}(15)$$ 
where
$$\frac{q}{2}=\frac{a^{3}}{27} - \frac{ab}{6}$$
$$ + \frac{c}{2},\hspace*{5cm}(8)$$
As the estimations show the period of oscillations approximately 
conform to the result of [7]. Now suppose that the nonlinear system's 
(4) parameters values are so that system (4) exhibits chaotic behavior 
and we will try to explore the possibility of chaos synchronization in 
the sense of Pecora and Carroll [10]. First consider the $x$ variable 
as a driver.Then variables $y, z$ form the replica response system 
(with the subscript "r"):
$$\frac{dy_{r}}{dt_{1}}=kx-xy_{r}-\beta y_{r}z_{r}$$
$$-\alpha y_{r}, \hspace*{3cm}(9)$$
$$\frac{dz_{r}}{dt_{1}}=g-\beta y_{r}z_{r}- \kappa z_{r}, $$
The eigenvalues of the Jacobian matrix of the system 
(9) satisfy the following equation:
$$\lambda^{2}+\lambda (x+\beta (y+z)+\alpha $$
$$+\kappa)+(\kappa+y\beta)(x+\alpha)$$
$$+\kappa\beta z=0, \hspace*{3cm}(10)$$
Here $x (t), y(t), z(t)$ are the solutions of the system (4).\\
As it follows from the physical meaning of $x, y, z$ they should be 
positive and bounded. The boundedness of the solutions of the 
nonlinear system (4) also follows from the Lorenz criteria (for more 
details, see [19]).
It means that the roots of equation (10) are negative, and therefore 
chaos synchronization is possible in the case of $x$ driving.
Similar calculations show that in the case of $y, z$ drivings chaos 
synchronization also could be realizable. Indeed, for the case of $y$ 
driving the eigenvalues of the Jacobian matrix of the corresponding 
replica subsystem are calculated from the relationship:
$$ (\lambda -\kappa - y\beta)(\lambda +k+1-y)=0,\hspace*{5cm}(11)$$
As the calculations show for the above given values of control 
parameter $g$ and other parameters $k>y$. It means the both 
$\lambda$'s are negative and therefore chaos synchronization is 
possible. Taking into account the above mentioned inequality $k>y$ one 
can easily establish that in the case of $z$ driving chaos 
synchronization also is possible, as under this inequality
the roots of equation 
$$\lambda^{2}+\lambda (x+\beta z+\alpha +k+1-y)+\beta z (k+1-y)$$
$$ + x+\alpha = 0, \hspace*{3cm}(12)$$
are also negative. Thus we have established that the replica approach 
is sufficient to perform chaos synchronization in the case of 
nonlinear system (4). As it was underlined above, the nonreplica 
approach to chaos synchronization could be used not only in the case 
of failure of replica approach, but also in the case of success by 
the replica approach to perform chaos synchronization. In the latter 
case one can use the nonreplica approach, say to make larger (in 
magnitude) the negative conditional Lyapunov exponents, as the very 
important quantity of chaos synchronization time 
inversely proportional to the largest conditional Lyapunov exponent. 
The application of the nonreplica approach (in our case) also could be 
justified in situations when conditions $k>y, k+1>y$ do not take 
place. (According to the definition of $k$ (see (3)) such a situation 
could be realizable by changing the values of $\alpha_{1},
N_{t}, c_{1}$). As it can be seen easily, then in the case of $y$ 
driving within the replica approach, one of $\lambda$'s become 
positive, and therefore chaos synchronization is impossible. Then to 
perform chaos synchronization we could recourse to the help of the 
nonreplica approach. As the calculations show under the above 
mentioned conditions $k<y, k+1<y$ the realizability of chaos 
synchronization in the case of $z$ driving is also under question. 
Here we will restrict ourselves to the nonreplica approach in the 
case of $z$ driving. The simplest nonreplica response system [15]
could be written in the following manner:
$$\frac{dx_{nr}}{dt_{1}}=g-(k+1)x_{nr} + x_{nr}y+\alpha y$$
$$ + s_{1} (y_{nr}-y)=A_{1} $$
$$\frac{dy_{nr}}{dt_{1}}=kx_{nr}-x_{nr}y-\beta yz_{nr}$$
$$-\alpha y +s_{2} (y_{nr}-y)=A_{2}, \hspace*{3cm}(13)$$
$$\frac{dz_{nr}}{dt_{1}}=g-\beta yz_{nr}- \kappa z_{nr}$$
$$ + s_{3}(y_{nr}-y)=A_{3}, $$
where $s_{1}, s_{2}, s_{3}$ are some arbitrary parameters (constants).
The eigenvalues of the Jacobian
$$J=\frac{\partial A}{\partial x},\hspace*{2cm} (14)$$ 
(where $A=(A_{1}, A_{2}, A_{3}), x=(x_{nr}, y_{nr}, z_{nr})$)
are to be found from the equation:
$$\lambda^{3} +\lambda^{2} (k+1-y+\kappa +\beta y -s_{2})$$
$$+ \lambda (k+1-y)(\kappa +\beta y-s_{2})-s_{2}(\kappa +\beta y)$$
$$-s_{1}(k-y)+\beta ys_{3})$$
$$+(k+1-y)(\beta ys_{3}-s_{2}(\kappa +\beta y)-s_{1}(k$$
$$-y)(\kappa+\beta y)=0,\hspace*{3cm} (15)$$
Our aim is to make negative the roots of this equations without 
explicit calculations. To achieve our goal we can use the following 
advantages: 1) As we have mentioned above the solutions of the 
nonlinear system is bounded; 2) the equation for $\lambda$ contains 
some arbitrary parameters $s_{1}, s_{2}, s_{3}$. In addition, there is 
no need to calculate these roots explicitly (in the case of higher 
dimensional systems the task is rather formidable) and investigate 
their dependence on the arbitrary constants with the aim of making the 
roots negative. For this purpose one can just use the Routh-Hurwitz 
criteria: the necessary and sufficient conditions for the equation 
(15) to have roots with negative real parts are : 
$$a_{1}>0, a_{3}>0, a_{1}a_{2}-a_{3}>0, \hspace*{4cm}(16)$$
where $a_{1}, a_{2}, a_{3}$ are coefficients of $\lambda^{2}, 
\lambda^{1}, \lambda^{0}$ in the equation (15), respectively.
If the nonlinear chaotic system's parameters value permits the 
condition $k>y$, then as it is clear from the equation for $\lambda$ , 
just the presence of larger negative $s_{2}$ is sufficient to make 
negative the roots of the equation (15). In other words, we can 
safely put $s_{1}=s_{3}=0$.If $k-y<0, k+1-y<0$ then larger negative 
$s_{2}$ and larger (in comparison with the absolute value of $s_{2}$) 
positive $s_{1}$ values with $s_{3}=0$ could make realizable chaos 
synchronization within the nonreplica approach. Without doubt, the 
presence of all the arbitrary constants $s_{1, 2, 3}$ in principle 
further alleviates the task of making negative the roots of the 
equation (15).\\
Thus in this paper we have shown that using the boundedness of the 
solutions of the some of the charge-carrier dynamical systems, 
nonreplica approach to chaos synchronization and Routh-Hurwitz 
criteria one can make negative the real parts of the conditional 
Lyapunov exponents without formidable, cubmersome and tedious 
numerical and analytical calculations.(For more detailed presentation 
of this approach to chaos control, see [20]).\\ 

\end{document}